\documentclass[journal=jacsat,manuscript=article]{achemso}
\usepackage[version=3]{mhchem} 

\author{Patrick T. Gemperline}
\affiliation{Department of Physics, Auburn University, Auburn, AL, USA}
\altaffiliation{These authors contributed equally to this work.}

\author{Arashdeep S. Thind}
\affiliation{Department of Physics, University of Illinois Chicago, Chicago, IL, USA}
\altaffiliation{These authors contributed equally to this work.}

\author{Chunli Tang}
\affiliation{Department of Electrical and Computer Engineering, Auburn University, Auburn, AL, USA}

\author{George E. Sterbinsky}
\affiliation{Advanced Photon Source, Argonne National Laboratory, Lemont, IL, USA}

\author{Boris Kiefer}
\affiliation{Department of Physics, New Mexico State University, Las Cruces, NM, USA}
\author{Wencan Jin} 
\affiliation{Department of Physics, Auburn University, Auburn, AL, USA}

\author{Robert F. Klie}
\affiliation{Department of Physics, University of Illinois Chicago, Chicago, IL, USA}
\email{rfklie@uic.edu}

\author{Ryan B. Comes}
\affiliation{Department of Physics, Auburn University, Auburn, AL, USA}
\alsoaffiliation{Department of Materials Science and Engineering, University of Delaware, Newark, DE, USA}
\email{comes@udel.edu}

\title[Strain Effects in SrHfO$_3$]{Strain Effects in SrHfO$_{3}$ Films Grown by Hybrid Molecular Beam Epitaxy}

\abbreviations{IR,NMR,UV}

\begin{document}








\begin{abstract}
Perovskite oxides hetero-structures are host to a large number of interesting phenomena such as ferroelectricity and 2D-superconductivity. Ferroelectric perovskite oxides have been of significant interest due to their possible use in MOSFETs and FRAM. SrHfO$_3$ (SHO) is a perovskite oxide with pseudo-cubic lattice parameter of 4.1~\AA$ $ that previous DFT calculations suggest can be stabilized in a ferroelectric P4mm phase, similar to STO, when stabilized with sufficient compressive strain. Additionally, it is insulating, possesses a large band gap, and a high dielectric constant, making it an ideal candidate for oxide electronic devices. In this work, SHO films were grown by hybrid molecular beam epitaxy with a tetrakis(ethylmethylamino)hafnium(IV) source on GdScO$_3$ and TbScO$_3$ substrates. Equilibrium and strained SHO phases were characterized using X-ray diffraction, X-ray absorption spectroscopy, and scanning transmission electron microscopy to determine the perovskite phase of the strained films, with the results compared to density functional theory models of phase stability versus strain. Contrary to past reports, we find that compressively-strained SrHfO$_3$ undergoes octahedral tilt distortions and most likely takes on the I4/mcm phase with the \textit{a\textsuperscript{0}a\textsuperscript{0}c\textsuperscript{-}} tilt pattern.
\end{abstract}

\section{Introduction}

    Complex metal perovskite oxides possess a wide range of electrical properties that have made them the focus of many works. They show promise as lead-free piezoelectrics, thermoelectrics, high $\kappa$ gate dielectrics in MOSFETs, etc. Some of the interesting phenomena are found at interfaces such as 2DEGs at LaAlO$_3$ (LAO)/SrTiO$_3$ (STO) interfaces~\cite{2DEG} and superconductivity at LAO/KTaO$_3$ interfaces~\cite{superconductor}. Perovskite oxides, such as STO and BaTiO$_3$, possess ferroelectric phases and have been investigated as possible materials for ferroelectric field effect transistors (FeFET) and ferroelectric random access memory (FRAM) devices~\cite{BTO-STO_FeFET, STO-Ferroelectric}. Good FeFET gate oxide candidates need to be ferroelectric, possess large band gaps, and have high dielectric constants. This has made Hf-based materials such as HfO$_2$ intriguing, due to the unusually weak ferroelectricity observed in the material and the large band gap and dielectric constant that are well known~\cite{mulaosmanovic2021ferroelectric}. For these reasons, demonstration of a hafnate material with a significant ferroelectric response has generated interest in the study of perovskite thin films.


    SrHfO$_3$ (SHO) is a perovskite oxide, similar to STO, where corner-sharing HfO$_6$ octahedra form the building blocks of the three-dimensional framework. The cuboctahedral cavities thus formed by these corner-connected HfO$_6$ octahedra are filled by Sr cations. SHO exhibits polymorphism and has various energetically competing phases. These phases correspond to space-group symmetries of Pnma, I4/mcm, P4/mbm, P4mm, Pm$\bar3$m, where Pnma is the ground state structure. SHO is a large band gap (6.1~eV)~\cite{SHO_Optical} perovskite oxide that is orthorhombic (Pmna)~\cite{High_Temp_Phases} in the bulk with a pseudocubic lattice constant of 4.08~\AA~\cite{SHO_Optical}. Temperature-dependent studies have found that SHO can also take on Cmcm, I4/mcm, and Pm$\bar3$m phases at higher temperatures~\cite{High_Temp_Phases}. Due to its high band gap and high dielectric constant (21)~\cite{Dielectric_Constant}, SHO has been investigated as a high-$\kappa$ gate dielectric for MOSFETs~\cite{Gate, ALD}. It has also been considered as a possible lead-free piezoelectric material~\cite{Piezoelectric,Bad_P4mm}. 
    
    Unlike other phases of SHO P4mm is non-centrosymmetric, which results primarily from the off-centering of Hf atoms with respect to the center of the HfO\textsubscript{6} octahedra. There have been several theoretical investigations into the possible existence of a  polar P4mm phase, with several predicting it does not exist~\cite{DFT,SHO_&_SZO_DFT}and others predicting its stability~\cite{Piezoelectric, Bad_P4mm}. If synthesized, these works predict the P4mm phase to be ferroelectric~\cite{DFT}, with a spontaneous polarization of 0.52~C/m$^2$ along the c axis~\cite{Piezoelectric}. In the last few years, there have been attempts to synthesize this P4mm phase with pulsed laser deposition~\cite{Bad_P4mm, PHO}. One study reported the successful synthesis of $\sim$35~nm thick P4mm SHO films on SrTiO$_3$ (STO) substrates using PLD and found evidence of ferroelectric behavior~\cite{Bad_P4mm}. However, another study synthesized SHO thin films on STO by PLD and found them to be cubic Pm$\bar3$m and paraelectric~\cite{PHO}. However, neither work was able to achieve coherently strained SHO films on STO, where the lattice mismatch is $\sim$-5\%. This indicates that studies exploring films with coherent compressive strain are still needed to understand the viability of a strain-induced polar phase.
    
    A key structural characteristic of a perovskite framework is for the BO\textsubscript{6} octahedra to undergo cooperative rotations or tilts. These octahedral tilts are denoted by Glazer’s notation~\cite{Glazer1,Glazer-2}. Using Glazer’s notation, an octahedral tilt pattern can be described as \textit{a}\textsuperscript{\textit{x}}\textit{b}\textsuperscript{y}\textit{c}\textsuperscript{\textit{z}}, where \textit{a}, \textit{b}, and \textit{c} correspond to the magnitude of the octahedral tilts along those respective crystallographic directions for a pseudocubic unit cell. If the magnitude of the octahedral tilts along two crystallographic directions is equal, then those tilts are denoted by the same symbol. The superscript (\textit{x}, \textit{y}, and \textit{z}) denotes the type of octahedral tilts, where (+) denotes in-phases octahedral tilts for the top and bottom neighboring octahedra, while (-) denotes out-of-phase octahedral tilts. The absence of octahedral tilts along a specific crystallographic direction is denoted by (0). The Pnma phase of SHO has an octahedral tilt pattern of \textit{a\textsuperscript{+}b\textsuperscript{-}b\textsuperscript{-}}, the same as for the GSO and TSO substrates. The tetragonal phases I4/mcm and P4/mbm have tilt patterns of \textit{a\textsuperscript{0}a\textsuperscript{0}c\textsuperscript{-}}and \textit{a\textsuperscript{0}a\textsuperscript{0}c\textsuperscript{+}}, respectively. The P4mm and the cubic Pm$\bar3$m phases do not exhibit octahedral tilts (\textit{a\textsuperscript{0}a\textsuperscript{0}a\textsuperscript{0}}).

    In addition to the possibility of possessing a ferroelectric phase, SHO also shows promise as a dielectric barrier layer in perovskite oxide heterostructures. Many perovskite oxides are unstable in the atmosphere or meta-stable for short periods of time, with exposure causing oxidation of the samples and sample degradation~\cite{Suresh}. This problem can be eliminated by adding a capping layer that buries the interfaces. However, this usually involves careful selection of a material based on its electronic and magnetic properties so that it will not affect the samples of interest~\cite{Living_Dead, SVO_Capping}. Density functional theory calculations for SHO have predicted that SHO's insulating behavior will prevent it from accepting donor electrons from any other transition metal perovskite oxide~\cite{Zhong}. This makes SHO an excellent candidate for capping atmospherically unstable and meta-stable perovskite oxides, as they would preserve the structure of the films while not changing the electrical properties of the films and interfaces. Theory also predicts that charge transfer through thin layers of SHO would still be possible, allowing for modulation doping of samples and the formation of 2DEGs at heterostructure interfaces~\cite{Zhong}. Hafnate materials have been employed as dielectrics to enable BaSnO$_{3}$ 2D electronic systems for possible device applications \cite{kim2017high, mahatara2022high}, where the lattice match is more favorable than other perovskite dielectrics. Development of MBE growth capabilities for SHO is important for interfacial electronic devices requiring a good dielectric barrier or capping layer.

    A key reason epitaxial SHO has not been synthesized using MBE is due to the refractory nature of Hf. Refractory metals have low vapor pressures even at temperatures approaching 2000\textdegree C. Thus, it is not practical to evaporate these elements in an effusion cell and an electron-beam evaporation source is needed instead. However, over the past few years, hMBE has made significant progress in repeatably and reliably utilizing such refractory metals to synthesize perovskite oxides~\cite{Suresh, Suresh_Ti, Brahlek, rimal2024advances}. hMBE achieves this by utilizing metal organic precursors commonly employed in atomic-layer deposition (ALD) as source material rather than pure metallic sources. Others have proposed the use of oxide sources in effusion cells that may evaporate as suboxide metal-oxide molecules and have proposed HfO as one possible source~\cite{adkison_suitability_2020}, but this has not yet been demonstrated and would still require evaporation from sources above 2000\textdegree C. Thus, metalorganic precursors with high vapor pressures at considerably lower temperatures offer the most promising avenue for consistent and well controlled deposition of Hf by MBE. Several Hf precursors are commercially available, but Tetrakis(ethylmethylamino)hafnium(IV) was chosen in this work due to its high vapor pressure, common use in atomic layer deposition, and thermal stability\cite{Hf_Precursor}.

    In this work, thin film samples of SHO were grown on STO, TbScO$_3$ (TSO, pseudocubic lattice parameter, $a_{pc}$ = 3.954 ~\AA~), and GdScO$_3$ (GSO, $a_{pc}$ = 3.963 ~\AA~) using hMBE. In situ RHEED was used to monitor the film surface during the growth process and \textit{in~vacuo} XPS~\cite{thapa2021probing} was used to investigate film composition and stoichiometry. HRXRD and RSM maps were used to determine film relaxation, while STEM was used to determine the film's atomic structure. Total energy calculations using DFT were carried out over a wide range of compressive and tensile strain to determine the thermodynamic stability of various phases of SHO. The electronic structure of the film and its dependence on the compressive strain was measured using XAS and compared with the electronic density of states calculations. The presence of polar distortions was investigated using SHG and STEM.

\section{First-principles Modeling}

\begin{figure}
    \centering
    \includegraphics[width=0.5\linewidth]{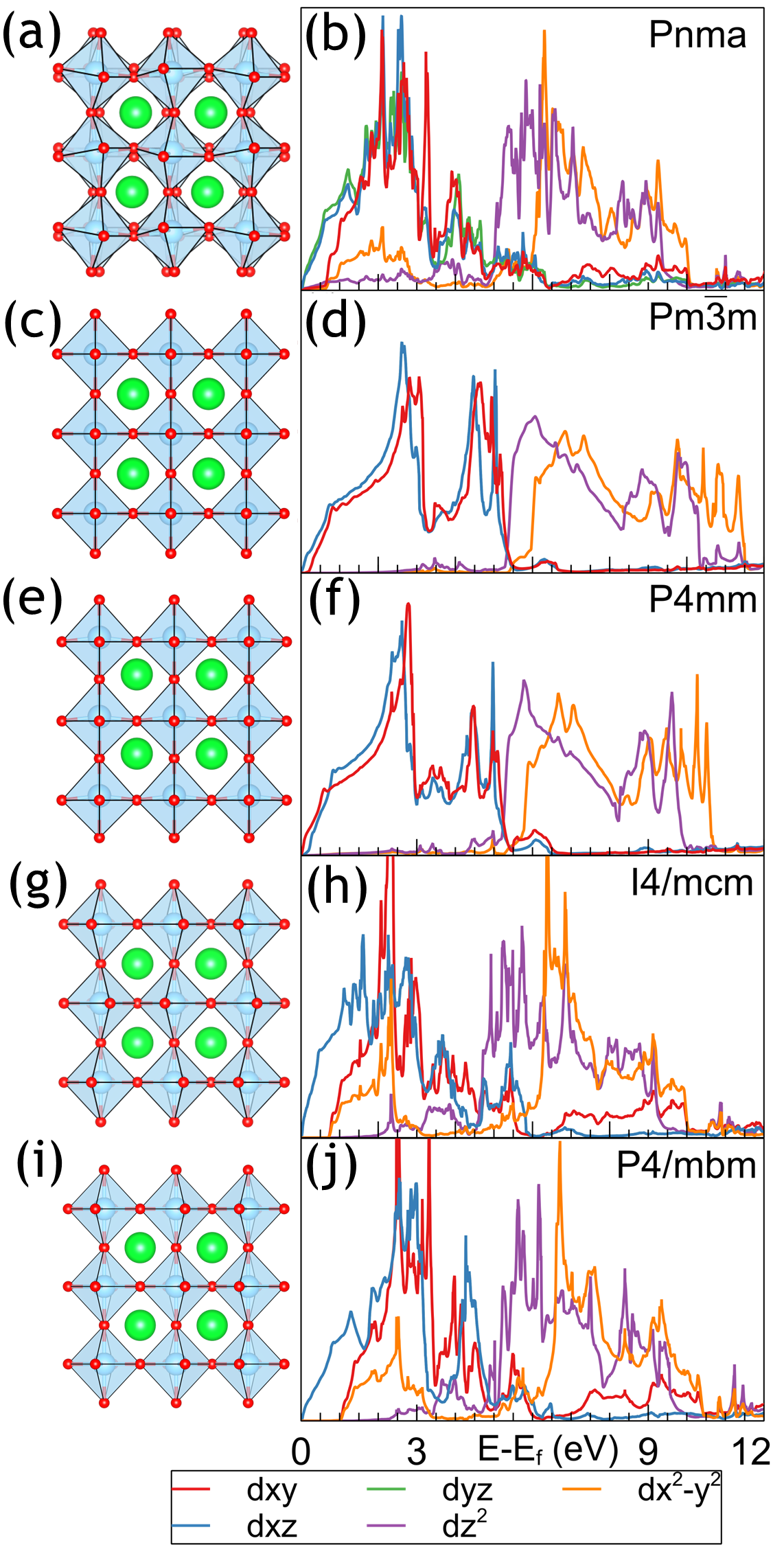}
    \caption{(a), (c), (e), (g), \& (i) Show the crystal structure of SHO for -3\% in-plane strain for the Pnma, Pm$\bar3$m, P4mm, I4/mcm, \& P4/mbm, respectively. (b), (d), (f), (h), \& (j) Show the eDOS results of our DFT results for the Pnma, Pm$\bar3$m, P4mm, I4/mcm, \& P4/mbm, respectively. The d$_{xz}$ and d$_{yz}$ states are degenerate for all phases but Pnma.} 
    \label{fig:DFT}
\end{figure}

\begin{figure}
    \centering
    \includegraphics{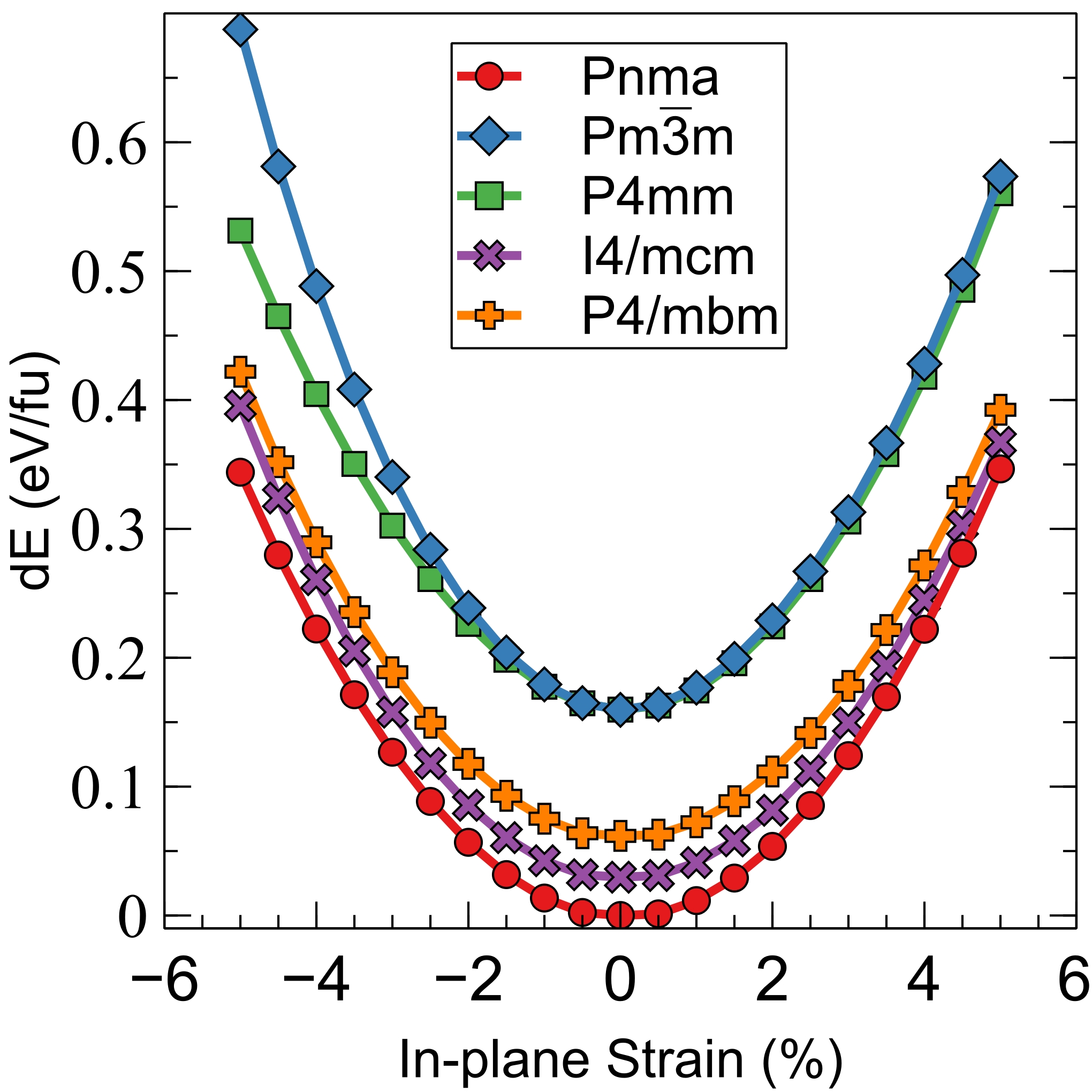}
    \caption{Plot of the change in energy from the ground state for the different SHO phases as a function of in-plane strain.} 
    \label{fig:E vs Strain}
\end{figure}

To predict phase-stability as a function of epitaxial strain, the thermodynamic ground state for 5~phases of SHO with with differing octahedral distortions was calculated using density functional theory (DFT). The computations were performed with the Vienna ab-initio Simulation Package~\cite{Boris1,Boris2} accounting for electronic exchange and correlations within the Perdew–Burke-Ernzerhof (PBE) parametrization of the Generalized-Gradient-Approximation (GGA)~\cite{Boris3}. Electrons are treated within the projector augmented-wave framework (PAW)~\cite{Boris4,Boris5}. We followed Material Project recommendations, and adopted a plane wave energy cutoff of E$_{cut}$~=~520~eV~\cite{Boris6} and a $\Gamma$-centered k-point grid with a k-spacing of 0.3~\AA$^{-1}$. We examined the phases: Pnma, I4/mcm, P4/mbm, Pm$\bar3$m, and P4mm. The equilibrium structures were subjected to in-plane strain ranging from -5\% to +5\% in steps of 0.5\% for each phase. The out-of-plane lattice parameter was calculated using the equation $\epsilon_{33} = \frac{-\nu}{1-\nu}(\epsilon_{11} + {\epsilon_{22}})$. The value of Poisson's ratio ($\nu$) of 0.3 for SHO was determined using lattice parameters obtained from XRD measurements of the samples described below. This value is consistent with the computed Poisson ratio of 0.25 for the Pm$\bar3$m phase. In order to determine the $e_g$ and $t_{2g}$ ordering, we computed the electronic density of states with energy bins of $<$~15~meV. The site projected eDOS d-orbital angular momentum channels in the d-orbital manifold were computed and normalized to arbitrary units (for phases with more than 1~Hf per~u.c.), providing phase resolved $t_{2g}$ ($d_{xy}$, $d_{xz}$, and $d_{yz}$) and $e_g$ ($d_{z^2}$ and $d_{x^2-y^2}$) ordering.\\ 
\indent Figure~\ref{fig:DFT} shows the crystal structure and electronic density of states for the Pnma, P4mm, Pm$\bar3$m, I4/mcm, and P4/mbm phases of SHO with -3\% compressive in plane strain. For P4mm we find a small lattice distortion of c/a=1.002, and that Hf is located above the equatorial plane of the HfO$_6$ octahedron by ~0.08~\AA, consistent with a ferroelectric phase. However we find that the P4mm phase is not the ground state. The energies for each phase as a function of in-plane strain are plotted in Figure~\ref{fig:E vs Strain}. The strain was calculated using $f=\frac{a_{substrate} - a_{film}}{a_{film}}$, where the pseudo cubic lattice constant and bulk lattice constant were used for the substrate and film respectively. From this figure we see that Pnma structured SHO is the thermodynamic ground state, consistent with previous computations~\cite{Bad_P4mm} and experiment~\cite{High_Temp_Phases}. Pnma remains the structural ground state for compressive and tensile strain in the range from -5\% to +5\%, covering the range of feasible strain values and experimental values. We find that phase stability decreases in the order Pnma (0~eV/fu) $<$ I4/mcm (30~meV/fu) $<$ P4/mbm (62~meV/fu) $<$ P4mm (160~meV/fu) $\sim$ Pm$\bar3$m (160~meV/fu) at 0\%~strain, consistent with previous work~\cite{Bad_P4mm}. We note, that the P4mm and Pm$\bar3$m equilibrium structures are energetically degenerate but structurally distinct. At -3\% compressive strain we find that phase stability decreases in the order Pnma (127~meV/fu) $<$ I4/mcm (158~meV/fu) $<$ P4/mbm (189~meV/fu) $<$ P4mm (303~meV/fu) $\sim$ Pm$\bar3$m (340~meV/fu).

\section{Experimental Methods}

    \subsection{Hybrid MBE}

    SHO films were grown in a (001) orientation on (110) oriented GSO and TSO with pseudocubic lattice parameters of 3.963~\AA\ and 3.954~\AA, respectively and on STO with lattice parameter 3.905~\AA. Substrates were obtained from MTI Crystal and sonicated in acetone and then isopropanol before being dried with nitrogen. GSO and TSO substrates were annealed for 6~hours at 1,100\textdegree C in a Lindberg Blue M tube furnace from Thermo Scientific. A Park Systems Atomic Force Microscope (AFM) was used to confirm this treatment resulted in a single termination with step edges $\sim$200~nm in width. All samples were grown in a Mantis MBE. The substrates were heated to 1,000\textdegree C over 1hr in an oxygen plasma with pressure 2~x~$10^{-6}$~Torr to prevent the surface of the substrate from reducing before growth. 
    
    Strontium ($99.99\%$, Sigma-Aldrich, USA) was supplied using a low-temperature effusion cell and the flux was calibrated with a quartz crystal micro-balance (QCM). An oxygen environment of 2~x~$10^{-6}$~Torr was used for calibration. Hf was supplied using the metal-organic precursor tetrakis(ethylmethylamino)hafnium (TEMAH, $99.99\%$, Sigma-Aldrich). TEMAH was stored in a bubbler and connected to the chamber by an ALD pneumatic valve (Swagelok 316L) and a heated gas injector (E-Science, USA). A capacitance manometer (Baratron) was attached to the gas line to monitor the Hf partial pressure. The temperature of the bubbler was maintained by a heating tape controlled by a PID controller. Hf~was calibrated by changing the bubbler temperature and observing the change in partial pressure. In order to reduce scattering of TEMAH and its constituent molecules during growth, the chamber shroud was cooled to -60\textdegree C using an SP~Scientific RC210 pump and Syltherm~XLT as coolant. 
    
    Samples were grown using co-deposition of Sr and Hf at 1,000\textdegree C in an oxygen plasma with chamber pressure 2~x~$10^{-6}$~Torr. Upon opening the Hf source, the chamber pressure increases to 7~x~$10^{-6}$~Torr and remains there for the duration of growth. After deposition, the samples were cooled from 1,000 to 400\textdegree C over 400~seconds in an oxygen plasma of 2~x~$10^{-6}$~Torr. Once the samples reached 200\textdegree C, they were transferred \textit{in~vacuo} to the XPS. During growth and cool down, \textit{in~situ} RHEED (Staib Instruments) was used to monitor the sample surface and growth quality. Videos of the RHEED were collected using Flashback Express Recorder. After the growth, principal component analysis and $k$-means clustering were performed on the RHEED using software previously developed by our group~\cite{Sydney}. 

    \subsection{Characterization}
    XPS spectra were collected for each sample using a PHI 5400 XPS with a base pressure of 8~x~$10^{-10}$~Torr. Low resolution spectra were taken over from 0 to 1400 Binding Energy(eV) with a pass energy of 178.95~eV and high resolution spectra were take over core level peaks with pass energies of 35.75~eV. The ratio of cations can be determined through XPS by comparing the areas under the fit curves and adjusting by the ratio of the relative sensitivity factors (RSF). In this way, the stoichiometry of each film was determined. The x-rays were generated by an Al K$\alpha$ source and an electron flood gun was used during data collection to prevent electron depletion since the film is not conducting. XPS spectra analysis and curve fitting was done using Casa XPS. 
    
    A Rigaku Smartlab XRD with a four-circle goniometer was used to collect 2$\theta$-$\omega$ scans over the (002) peak of each sample. This system utilizes the Cu~ K$\alpha_{1}$ line isolated with a double bounce Ge (220) monochromator. Reciprocal space maps were captured using a 2D detector and were used to determine film strain.

    Hafnium $L_3$-edge X-ray absorption spectroscopy was carried out at beamline 20-BM of the Advanced Photon Source at Argonne National Laboratory. The incident x-ray energy was controlled by a Si(111) double crystal monochromator, and the x-ray beam was focused by a Pt/alumina bilayer coated toroidal mirror. The Hf~$L\alpha$ partial fluorescence yield was collected using a seven element Ge solid state detector. All samples were spun about an axis normal to the film surface during measurement to mitigate Bragg peaks. Angles of incidence were maintained under ten degrees, and the x-ray polarization was oriented either in-plane or out-of-plane. 
    
    Rotational anisotropy second harmonic generation measurements were performed using an ultrafast light source with 800~nm wavelength, 50~fs pulse duration, and 200~kHz repetition rate. In the normal incidence geometry, the incident and reflected light were fixed as p or s polarization, and the reflected SHG intensity is recorded as a function of the azimuthal angle $\phi$ between the scattering plane (electric polarization) and the in-plane crystalline axis. The incident light was focused onto a 50~$\mu$m diameter spot on the sample with a fluence of $\sim$0.25~mJ/cm$^2$. The second harmonic signal is collected by a single photon counting EMCCD camera. 

    \subsection{Electron Microscopy}
    The cross-section lift-out samples of SrHfO\textsubscript{3}/TbScO\textsubscript{3} (SHO-TSO) and SrHfO\textsubscript{3}/GdScO\textsubscript{3} (SHO-GSO) films were prepared using a Thermo Fischer Scientific Helios 5~CX focused-ion beam (FIB)/ scanning electron microscope (SEM) DualBeam system at University of Illinois Chicago. Samples were prepared such that the SHO-TSO sample would have a zone axis of [010] for the orthorhombic TSO substrate, while the SHO-GSO sample had a [10$\bar1$] zone axis for the GSO substrate. This allowed for the observation of cation sublattice and octahedral tilts along both axes to determine the space group of the SHO films under strain. The SHO-TSO and SHO-GSO thin films were sputter-coated with a 10~nm thick layer of Pt/Pd to avoid charging during FIB lamellae preparation. To protect against the ion-beam induced surface damage to the SHO films, a protective coating of W was deposited. The final lamellae thinning was performed using a 1~kV ion beam energy to minimize the amorphization of the lamellae cross-section and to obtain electron-transparent samples.
    
    Scanning transmission electron microscopy (STEM) experiments were carried out at University of Illinois Chicago using an aberration-corrected JEOL JEM-ARM200CF microscope. The microscope is equipped with a cold-field emission gun and a CEOS aberration corrector, and was operated at 200~kV. The electron energy loss spectroscopy (EELS) experiments were performed using a dual-range Gatan Continuum spectrometer. A spectrometer entrance aperture of 5~mm was used, resulting in a collection semi-angle of 53.4~mrad. A dispersion of 0.75 eV per channel was used to acquire the core-loss edges in the low-loss (Sc~L and O~K) and high-loss (Tb~M, Gd~M, Hf~M, and Sr~L) energy range with an acquisition time of 0.25~seconds per spectrum. The background signal before core-loss edges was modeled using a power law. To improve EELS signal quality, principal component analysis (PCA) was performed to remove random noise components. An Oxford X-Max 100TLE windowless silicon drift detector was used to perform energy-dispersive X-ray spectroscopy (EDS). 
    
    A probe convergence semi-angle of 30~mrad was used to perform atomic-resolution high-angle annular dark-field (HAADF), low-angle annular dark-field (LAADF) and annular bright-field (ABF) imaging. The collection angles used for HAADF, LAADF and ABF imaging were set to 90~mrad to 370~mrad, 40~mrad to 160~mrad, and 11~mrad to 23~mrad, respectively. The atomic-resolution images were acquired sequentially (10-15~frames). The images were subsequently aligned and integrated to improve signal-to-noise ratio. For bond-distance analysis, the atomic positions in the HAADF images were initially estimated using relative intensity. Subsequently, these atomic positions were refined using 2D Gaussian fitting.
    
    Four-dimensional scanning transmission electron microscopy (4D-STEM) experiments were carried out using a Gatan ClearView CMOS detector. The microscope was operated at 200~kV with a probe convergence semi-angle of 2~mrad. The 4D-STEM datasets were acquired with a pixel size of 0.8~nm to 1~nm and the diffraction patterns were collected using an exposure time of 50~ms to 75~ms. A combination of hardware and software binning was used to limit the size of each diffraction pattern to 256~×~256 pixels. The 4D-STEM data analysis was performed using the py4DSTEM python package~\cite{4D_STEM}.

\section{Experimental Results}

    \subsection{Film Synthesis and Characterization}

\begin{figure*}
            
            \centering
            \includegraphics{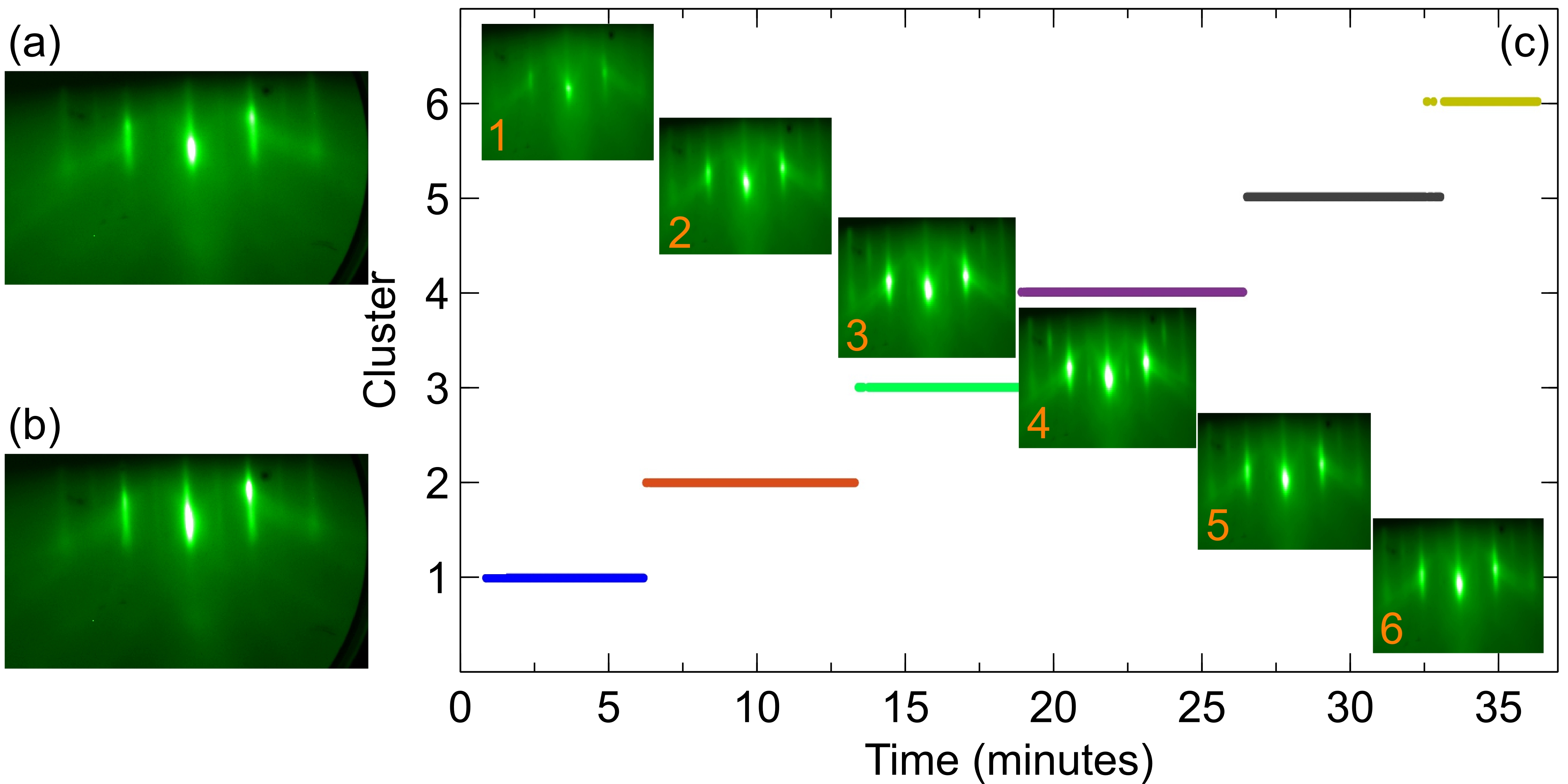}
            \caption{(a) RHEED image of SHO-TSO film after growth at 1000\textdegree C. (b) RHEED image of SHO-TSO film after growth at room temperature. (c) $k$-means clustering of SHO-TSO film. The mean RHEED image for each group is vertically aligned with the group and have the group number inlaid.}
            \label{fig:Kmeans}
        \end{figure*}

        
        Figure \ref{fig:Kmeans}(a) shows the high temperature RHEED image of SHO-TSO along the [110] azimuth immediately following the conclusion of growth, while Figure \ref{fig:Kmeans}(b) shows the ambient temperature RHEED image. The bright spots and streaks with clear Kikuchi bands indicate a smooth and highly crystalline surface. The intermediate streaks indicate a weak surface reconstruction is present. From the $k$-means clustering in Figure~\ref{fig:Kmeans}(c), we can see how the surface evolves throughout the 37~minute long growth of SHO-TSO. Each cluster is constructed by minimizing the difference between each frame and the mean frame of the cluster. As a result, each frame in the cluster is most similar to the mean of that cluster than the mean of any other cluster. Cluster 1 encompasses the TSO RHEED pattern and shows a decrease in intensity and softening Kikuchi bands, which is may be attributed to the polar-non-polar interface and disorder in the early stages of growth. The intensity increases in clusters 2,~3,~\&~4, with 3~\&~4 showing a weak surface reconstruction between the primary spots. This indicates the surface was improving and was highly crystalline and well ordered. The presence of Kikuchi bands in Figure \ref{fig:Kmeans}(b) indicate the surface retains its fairly smooth and retains high crystallinity even after cooling down to ambient temperature. 

\begin{figure*}
            \centering
            \includegraphics{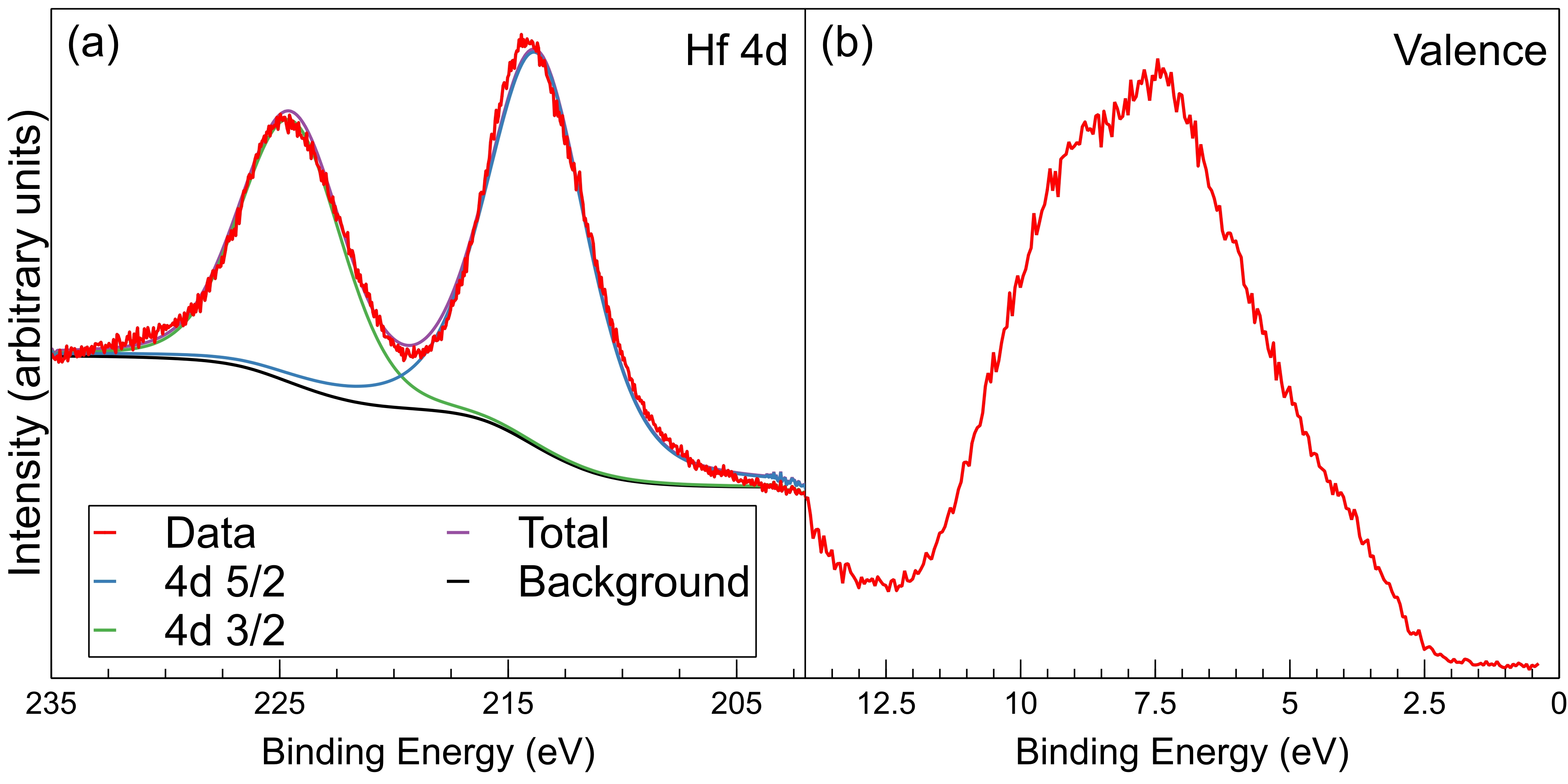}
            \caption{(a) XPS spectra of the Hf 4d peak from SHO-TSO film. (b) XPS spectra of the valence band of SHO-TSO film.} 
            \label{fig:XPS}
        \end{figure*}

        
        Figure~\ref{fig:XPS}(a) shows the Hf 4d peak convolutions for SHO-TSO and Figure \ref{fig:XPS}(b) shows the valence band spectra collected at the same time. Peak fits for O~1s and Sr~3d can be found in the supplement. From the Hf~4d and Sr~3d peaks, the cation ratio of the SHO-TSO sample was calculated to be approximately 1:1, indicating the film is nearly stoichiometric. Since strontium has a significantly lower atomic mass than terbium, RBS would not provide accurate stoichiometry. 
        
\begin{figure*}
            \centering
            \includegraphics{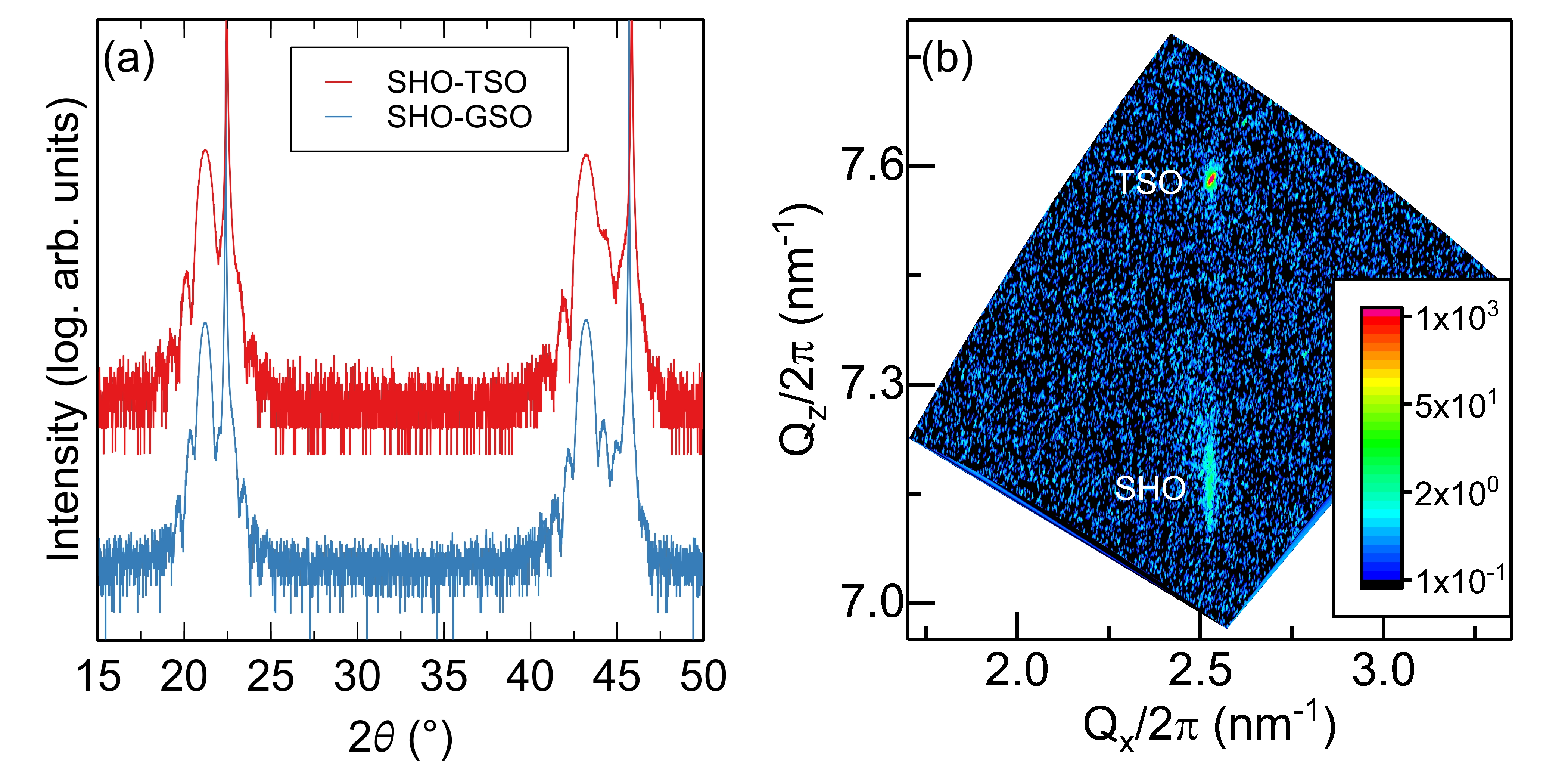}
            \caption{(a) HRXRD of SHO-TSO and SHO-GSO samples. (b) RSM of SHO-TSO film showing the SHO film is epitaxially strained with TSO.} 
            \label{fig:XRD}
        \end{figure*}

        Figure~\ref{fig:XRD}(a) shows the HRXRD spectra of SHO films grown on TSO and GSO substrates. In both samples we observe a film peak at lower 2$\theta$ values than the TSO and GSO substrate peaks indicating the films have larger out of plane lattice constants. There are no peaks indicating other phases are present. Figure~\ref{fig:XRD}(b) shows the reciprocal space map for the 103 pseudo-cubic peak for the SHO-TSO film. The sharpness of the SHO film peak in the $Q_x/2\pi$ axis indicates the samples possess a very small range of in plane lattice constants. Since the SHO film spot and the TSO substrate spot are aligned at the same value on the $Q_x/2\pi$ axis, we know the film possesses the same in-plane lattice constant as the substrate and thus has a compressive strain of $\sim$-3\%.

    \subsection{Characterization of Film Space Group}
    
        \subsubsection{Second Harmonic Generation}
        
\begin{figure*}
            \centering
            \includegraphics{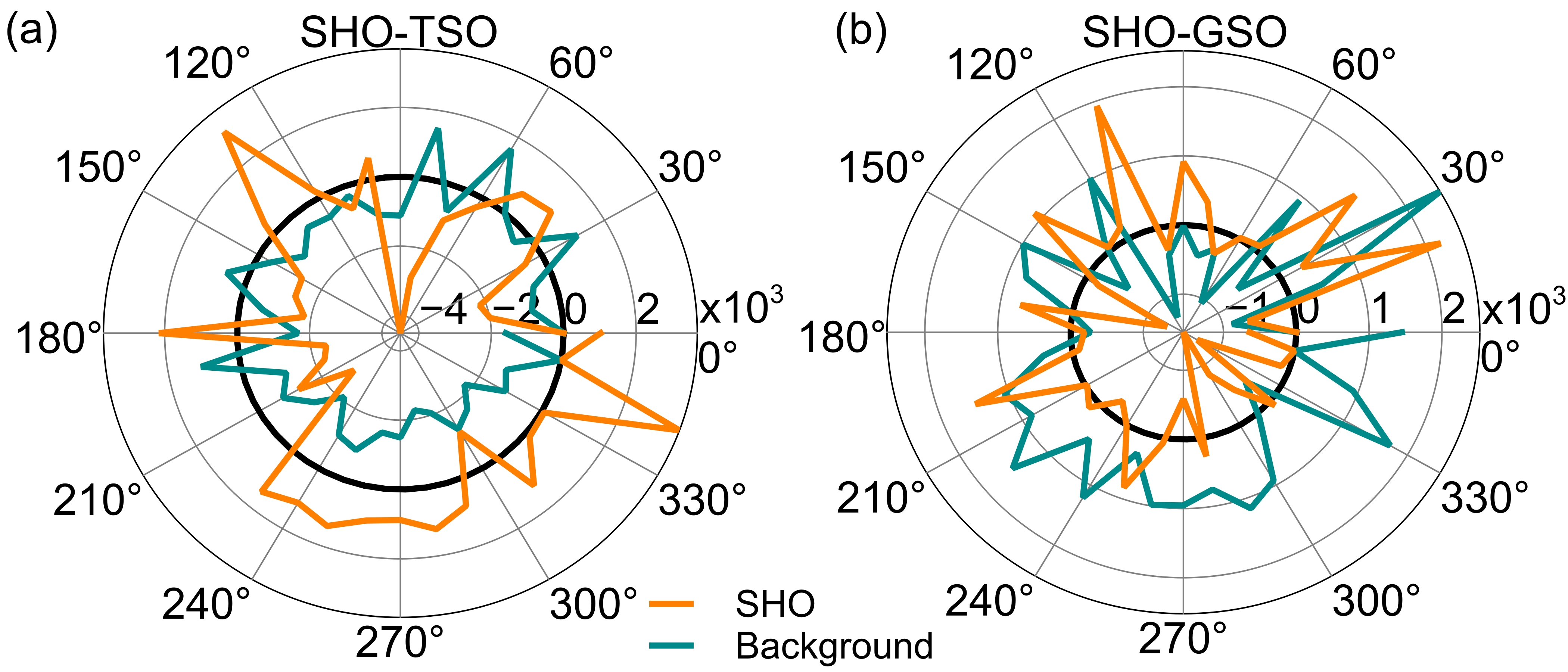}
            \caption{ SHG measurements of SHO-TSO (a) and SHO-GSO(b). The lack of significant signal shows neither film is centrosymmetric.} 
            \label{fig:SHG}
        \end{figure*}

        Figure \ref{fig:SHG} shows the results of SHG measurements for SHO-TSO and SHO-GSO samples. The results show no sign of a SHG response indicating that the SHO films are centrosymmetric. Of the 5 phases expected for SHO, only the P4mm phase is polar and non-centrosymmetric. Thus, this result is consistent with the Pnma, Pm$\bar3$m, I4/mcm, and P4/mbm phases of SHO and inconsistent with the emergence of a strain-stabilized P4mm phase. The lack of SHG response also supports the absence of a ferroelectric distortion in strained SHO predicted by previous studies~\cite{Piezoelectric, DFT}. 

        The absence of a P4mm phase is consistent with the DFT modeling presented above which finds that there is a 160~meV/fu difference between the P4mm and ground state Pnma phases. The absence of a polar phase matches the results found by Acharya et al.~\cite{PHO}, which observed a Pm$\bar3$m phase in SHO films grown on STO by PLD. However, other work~\cite{Bad_P4mm} found that 35~nm thick SHO-STO films grown by PLD possessed a P4mm phase and observed a SHG response. The samples synthesized in these works shown by RSM appeared to be relaxed, as compared to the films presented here which are compressively strained to the GSO and TSO substrates. Nonetheless, we do not see evidence of a strain-induced polar phase, consistent with the DFT predictions of phase stability presented above.
        
        \subsubsection{X-ray absorption spectroscopy}
    
\begin{figure*}
            \centering
            \includegraphics{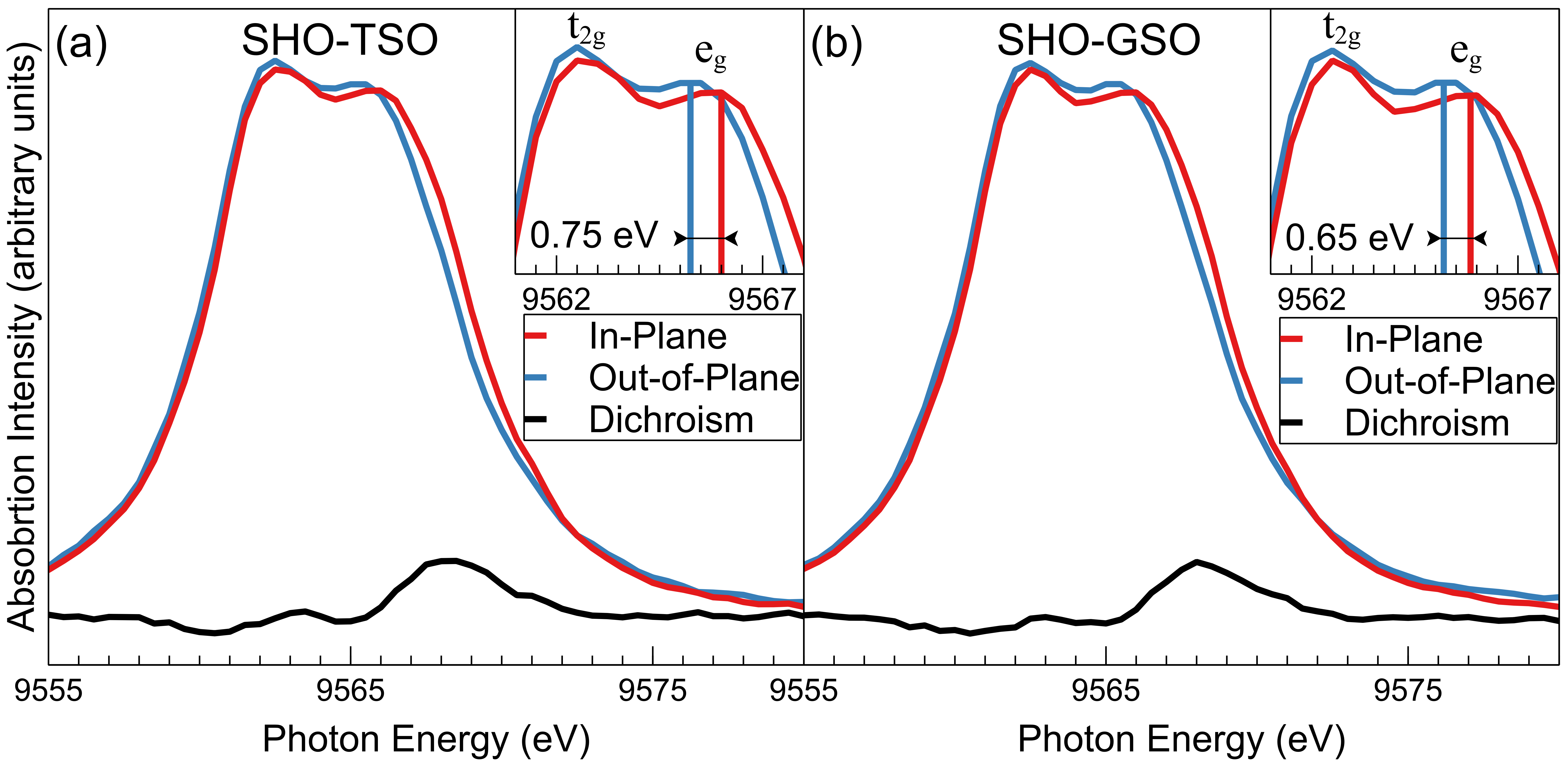}
            \caption{XAS data for SHO-TSO (a) and SHO-GSO (b). Insets show the splitting between in-plane and out-of-plane signal for the e$_g$ band.}
            \label{fig:XANES}
        \end{figure*}

        Figure~\ref{fig:XANES} (a) \& (b) depict the Hf L$_3$-edge for an SHO-TSO film and SHO-GSO film respectively. These measurements were performed to probe splitting of the energy levels for unoccupied Hf~$5d$ states under strain and to the best of our knowledge are the first reported spectra Hf~L$_3$ spectra for perovskite hafnates. The absorption spectra show clear splitting of the t$_{2g}$ and e$_g$ levels in the absorption, which generally has not been observed in measurements of HfO$_2$ samples previously~\cite{cho2010structural}. The X-ray linear dichroism (XLD)is calculated by XLD$=I_{ab}-I_{c}$. In both samples, the t$_{2g}$ peaks are nearly degenerate between the in-plane and out-of-plane response, with the leading edge of the out-of-plane $\sim$100~meV. However, the samples exhibit a significantly larger splitting of $\sim$700~meV between the the in-plane and out-of-plane spectra at the e$_{g}$ peaks. 
        
        As result of crystal field splitting, the Hf unoccupied $5d$-orbitals are broken into the t$_{2g}$ and e$_{g}$ bands where $d_{xz} = d_{yz} = d_{xy} < d_{z^{2}} = d_{x^{2}-y^{2}}$. As observed from the RSM and XRD, these samples are compressively strained in-plane due to the smaller substrate lattice parameter and possess increased c-lattice parameters as compared to the bulk. This elastic symmetry-breaking should induce additional crystal field splitting that breaks the degeneracy of the t$_{2g}$ and e$_{g}$ bands resulting in $d_{xz} = d_{yz} < d_{xy} < d_{z^{2}} < d_{x^{2}-y^{2}}$. This can be observed in the SHO samples by the presence of the weak dichroism in the leading edge and t$_{2g}$ band peak and a more substantial dichroism at the e$_g$ peak. This result is consistent with past study on BTO-DyScO$_{3}$ (DSO) films, which observed a splitting of $\approx$1~eV was observed in the e$_{g}$ band peak, similar to the splitting observed here~\cite{XAS}. While the SHO films exhibit a slight dichroism in the t$_{2g}$ band, the BTO-DSO films observed a triply degenerate t$_{2g}$ band. This degeneracy in the t$_{2g}$ was attributed to the presence of polar distortions, which offset the effect of the crystal field splitting in BTO \cite{XAS}.
        
        To compare our results with theory, we applied a Gaussian convolution to the eDOS predicted by our DFT calculations to match the XAS instrumental resolution. The resulting convolutions for all 5~phases at -3\% compressive strain can be found in the supplement, with the Pnma phase shown in Figure~\ref{fig:Convolution}. As can bee seen in Figure~S7, all phases except the P4mm exhibit a dichroism on the L$_3$-edge. This dichroism is also exhibited by both SHO samples in Figure~\ref{fig:XANES}. This result is in agreement with the SHG measurements and effectively rule out the presence of the P4mm phase. As can be seen in Figure~S7, the Pm$\bar3$m phase possesses an additional feature on the high energy side of the L-edge, not present in the other modeled phases or the SHO XANES spectra. In Figure~\ref{fig:Convolution}~(b-c) the dichroisms and their derivatives for each phase possess similar features, with exception of the additional peak on the Pm$\bar3$m phase at 10~eV. These features along with the significant decrease in phase stability predicted for by DFT, rule out the presence of the Pm$\bar3$m phase. Determination of the stable octahedral tilt pattern from XAS was not possible due to the very similar XLD predictions for the Pnma, I4/mcm, and P4/mbm phases.

        \begin{figure*}
            \centering
            \includegraphics{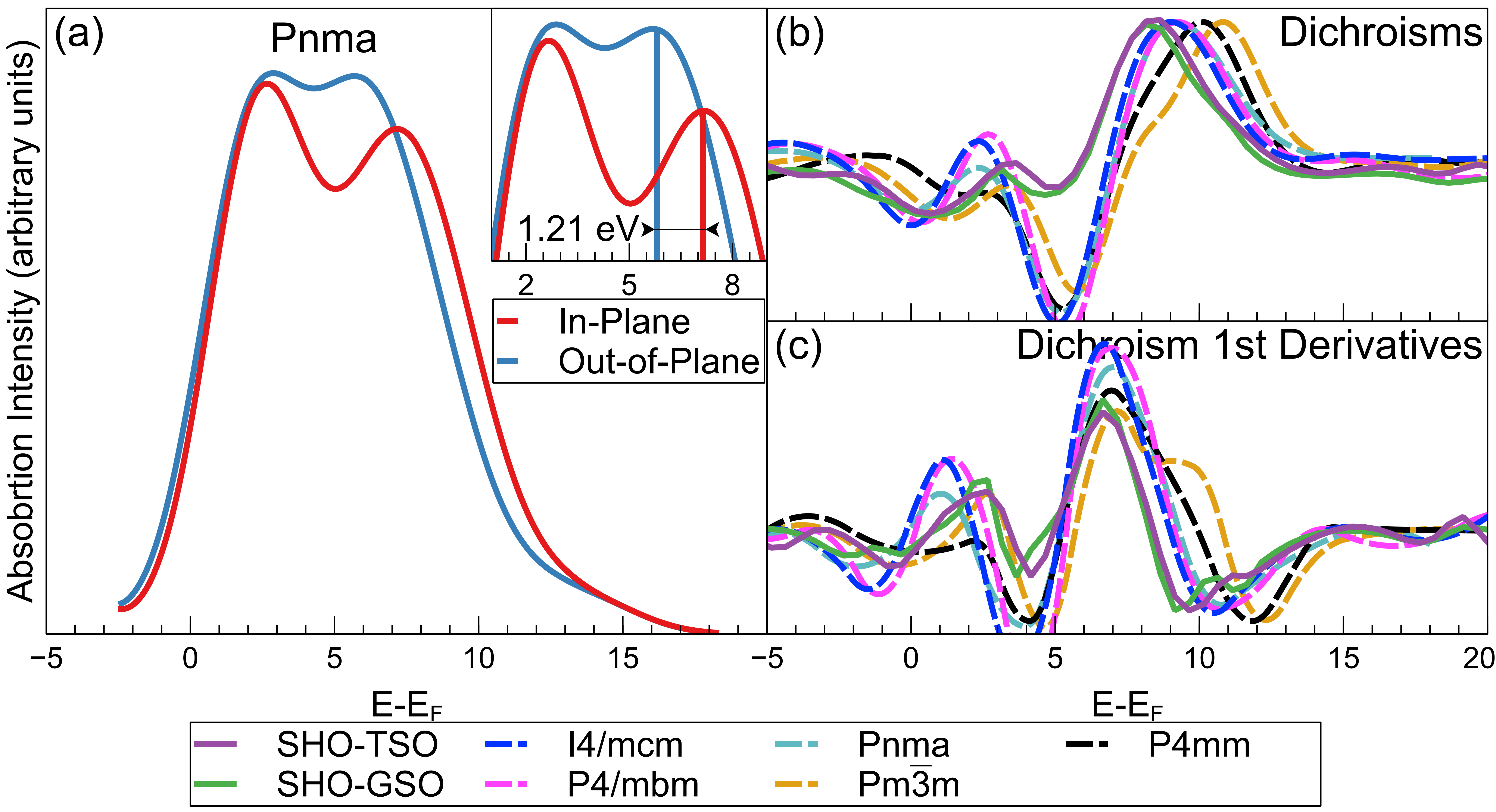}
            \caption{(a) The Gaussian convolution of the eDOS from DFT results for the Pnma phase of SHO. (b) \& (c) The dichroism and it's derivative, respectively, for the XAS data of SHO-TSO, SHO-GSO, and the 5~phases of SHO modeled by DFT with -3\% in-plane strain.}
            \label{fig:Convolution}
        \end{figure*}

        \subsubsection{Scanning Transmission Electron Microscopy}
        
\begin{figure}
            \centering
            \includegraphics[width=1\linewidth]{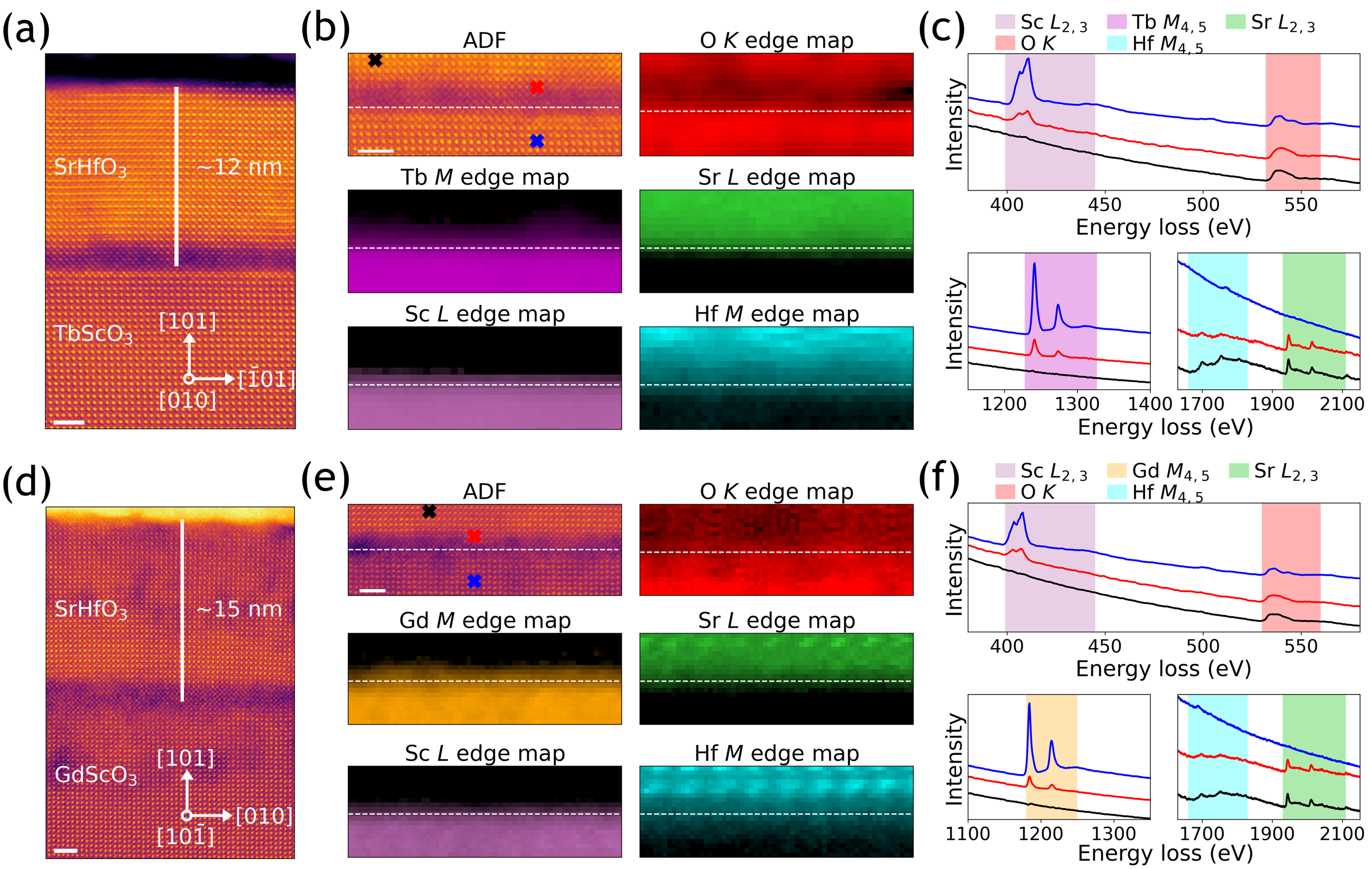}
            \caption{(a) Atomic resolution HAADF image showing the cross-section of the SHO-TSO film along [010]. (b) ADF image along with EELS chemical maps of Tb~\textit{M}, Sc~\textit{L}, O~\textit{K}, Sr \textit{L}, and Hf~\textit{M} edges across the SHO-TSO interface showing the chemical distribution of elements. (c) Extracted EEL spectra for Tb~\textit{M}, Sc~\textit{L}, O~\textit{K}, Sr~\textit{L} and Hf~\textit{M} edges. The EEL spectra were extracted from the probe positions marked with the same color in the ADF image shown in (b). (d) Atomic resolution HAADF image showing the cross-section of the SHO-GSO film along [10$\bar1$]. (e) ADF image along with EELS chemical maps of Gd~\textit{M}, Sc~\textit{L}, O~\textit{K}, Sr~\textit{L}, and Hf~\textit{M} edges across the SHO-GSO interface showing the chemical distribution of elements. (f) Extracted EEL spectra for Gd~\textit{M}, Sc~\textit{L}, O~\textit{K}, Sr~\textit{L} and Hf~\textit{M} edges. The EEL spectra were extracted from the probe positions marked with the same color in the ADF image shown in (e). Scale bars in (a), (b), (d) and (e) correspond to 2~nm.}
            \label{fig:STEM-EELS}
        \end{figure}

        To investigate the atomic and chemical structure of the SHO-TSO and SHO-GSO films along with the interfacial strain effects in the samples, we have performed STEM imaging experiments combined with EELS and EDS. Figure~\ref{fig:STEM-EELS} shows the STEM-EELS results for the SHO-TSO and SHO-GSO films. Figure~\ref{fig:STEM-EELS}(a) and \ref{fig:STEM-EELS}(d) show the atomic resolution HAADF images of the SHO-TSO[010] and SHO-GSO[10$\bar1$] films along with their respective crystallographic orientations. We have performed EELS experiments to understand the chemical distribution of elements across the cross-section of the SHO-TSO and SHO-GSO. The Sr~\textit{L}, Hf~\textit{M}, Tb~\textit{M}, Gd~\textit{M}, Sc~\textit{L}, and O~\textit{K} edge maps are shown in Figures~\ref{fig:STEM-EELS}(b) and~\ref{fig:STEM-EELS}(e) for the SHO-TSO and SHO-GSO thin film samples, respectively. We observe the interface to be chemically diffused for both SHO-TSO and SHO-GSO samples. For instance, on comparing the EELS chemical maps for Tb~\textit{M} and Hf~\textit{M} edges (SHO-TSO) as well as Gd~\textit{M} and~Hf \textit{M} edges (SHO-GSO), we observe the interface is not chemically sharp but diffused. There is formation of a chemically intermixed layer, which is about $\approx$2-4 unit cells (u.c.) thick. To emphasize this point, the extracted EEL spectra for different regions of the film and substrate of the SHO-TSO and SHO-GSO samples are shown in Figures~\ref{fig:STEM-EELS}(c) and~\ref{fig:STEM-EELS}(f), respectively. We ascribe this chemical heterogeneity at the interface to non-uniform morphology and surface terminations of the TSO and GSO substrates. Moreover, we also observe the SHO-TSO and SHO-GSO interfaces to be oxygen deficient, as shown in O~K edge maps in Figures~\ref{fig:STEM-EELS}(b) and~\ref{fig:STEM-EELS}(e), respectively. The impact of the chemical heterogeneities at the surface of TSO and GSO substrates on the atomic structure of the SHO films is discussed later in the manuscript. Additionally, the EDS chemical maps for Sr~\textit{K$\alpha$}, Hf~\textit{L$\alpha$}, Tb~\textit{L$\alpha$}, Gd~\textit{L$\alpha$}, Sc~\textit{K$\alpha$}, and O~\textit{K$\alpha$} edges are shown in the supplementary information Figure S14.        

        Our DFT calculations (see Figure \ref{fig:E vs Strain}) show that SHO has various energetically competing phases for a range of compressive strains. These phases correspond to Pnma, I4/mcm, P4/mbm, P4mm, Pm$\bar3$m, where Pnma is the calculated ground state for a wide range of compressive and tensile strain as shown in Figure~\ref{fig:E vs Strain}. The atomic models of the SHO with in-plane compressive strain of 3\% are shown in Figure~S1. Both TSO and GSO also adopt the same space-group symmetry of Pnma as the calculated ground state of SHO. The atomic models showing the TSO and GSO substrate orientations used for SHO film growth are shown in Figure~S2.
        
\begin{figure}
          \centering
          \includegraphics[width=0.8\linewidth]{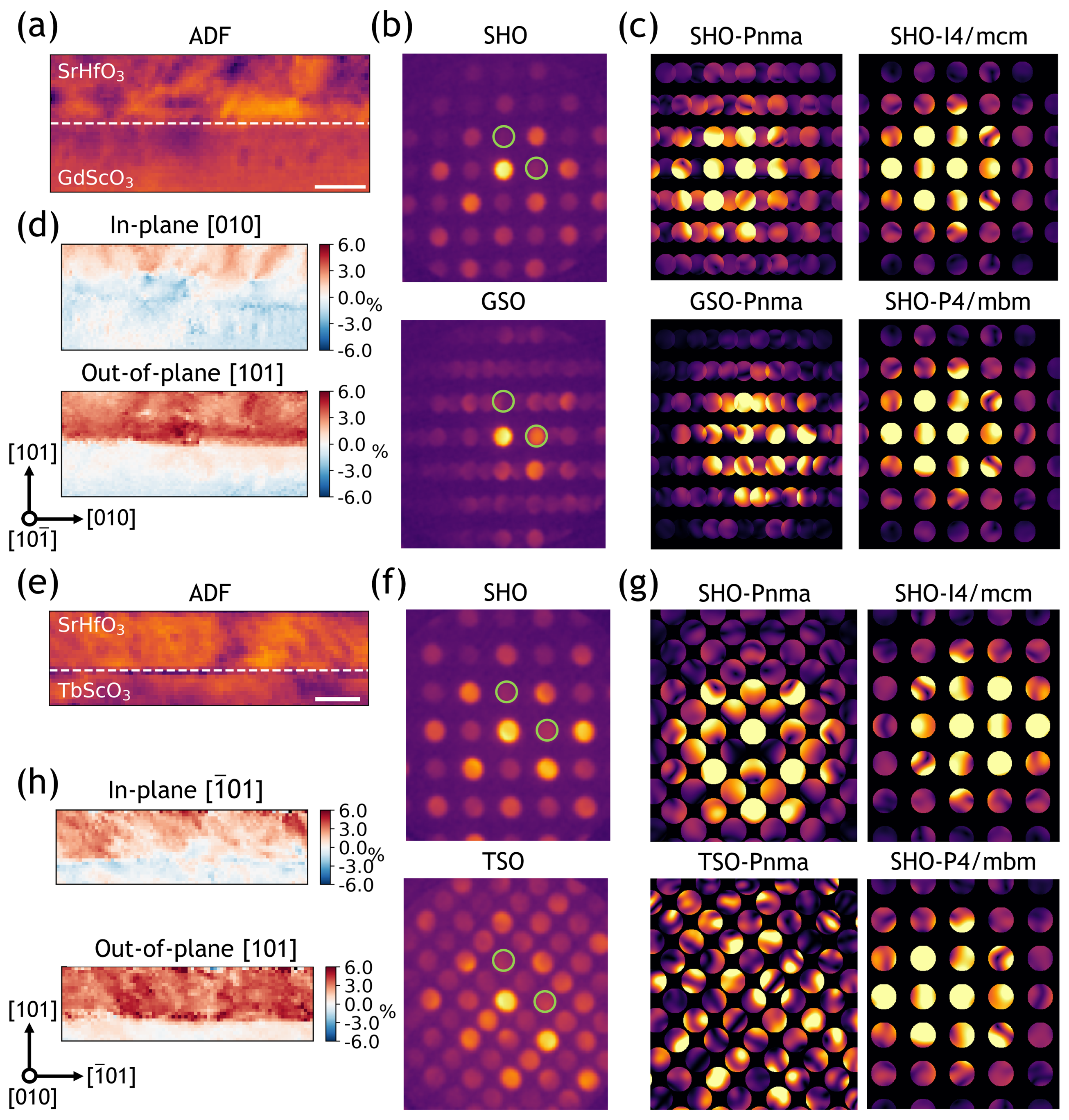}
          \caption{ADF image simultaneously acquired during 4D-STEM data collection for (a) SHO-GSO and (e) SHO-TSO samples. The dashed white lines mark the SHO-GSO/TSO interface. CBED patterns for (b) SHO-GSO and (f) SHO-TSO integrated over the area of the film and substrate respectively. Simulated CBED patterns for (c) SHO-GSO and (g) SHO-TSO. The crystallographic orientations corresponding to the experimental and simulated data are provided in (d) and (h) for SHO-GSO and SHO-TSO samples respectively. Strain maps for (d) SHO-GSO and (h) SHO-TSO samples, where strain is calculated with respect to the substrate. Bragg disks used for strain mapping are marked as green circles in (b) and (f) for SHO-GSO and SHO-TSO respectively. The CBED patterns have been rotated post acquisition to match the real-space substrate orientation. Scale bars in (a) and (e) correspond to 10~nm.}
          \label{fig:4D-STEM}
        \end{figure}

        To understand the experimental epitaxial relationships between SHO-GSO/SHO-TSO, and evaluate the microstructure of the SHO film, we have performed 4D-STEM experiments. 4D-STEM provides a unique capability of elucidating the space-group symmetry of the sample in the reciprocal space with a sufficient real-space resolution to clearly distinguish between the film and the substrate. Figure~\ref{fig:4D-STEM} shows the 4D-STEM results for SHO-GSO[10$\bar1$] and SHO-TSO[010]. Figures~\ref{fig:4D-STEM}(a) and~\ref{fig:4D-STEM}(e) show the ADF images simultaneously acquired during the 4D-STEM dataset collection for SHO-GSO and SHO-TSO respectively. The extracted convergent-beam electron diffraction (CBED) patterns for the SHO film and GSO/TSO substrates are shown in Figures~\ref{fig:4D-STEM}(b) and~\ref{fig:4D-STEM}(f). Figures \ref{fig:4D-STEM}(c) and~\ref{fig:4D-STEM}(g) show the simulated CBED patterns for the SHO-Pnma, SHO-I4/mcm, SHO-P4/mbm and GSO-Pnma/TSO-Pnma phases, along the GSO[10$\bar1$] and TSO[010] respectively. By comparing the CBED patterns for the SHO film and the corresponding substrates (GSO and TSO), we can clearly see that the SHO film does not adopt the expected Pnma phase. For the SHO film, we find that the simulated CBED patterns for SHO-I4mcm and SHO-P4mbm show a good match with the experimental CBED patterns along both GSO[10$\bar1$] and TSO[010] crystallographic orientations. Moreover, simulated kinematical diffraction patterns for a parallel electron beam condition are shown in Figure~S16 for all potential SHO phases along with the GSO and TSO substrates. 
        
\begin{figure}
            \centering
            \includegraphics[width=1\linewidth]{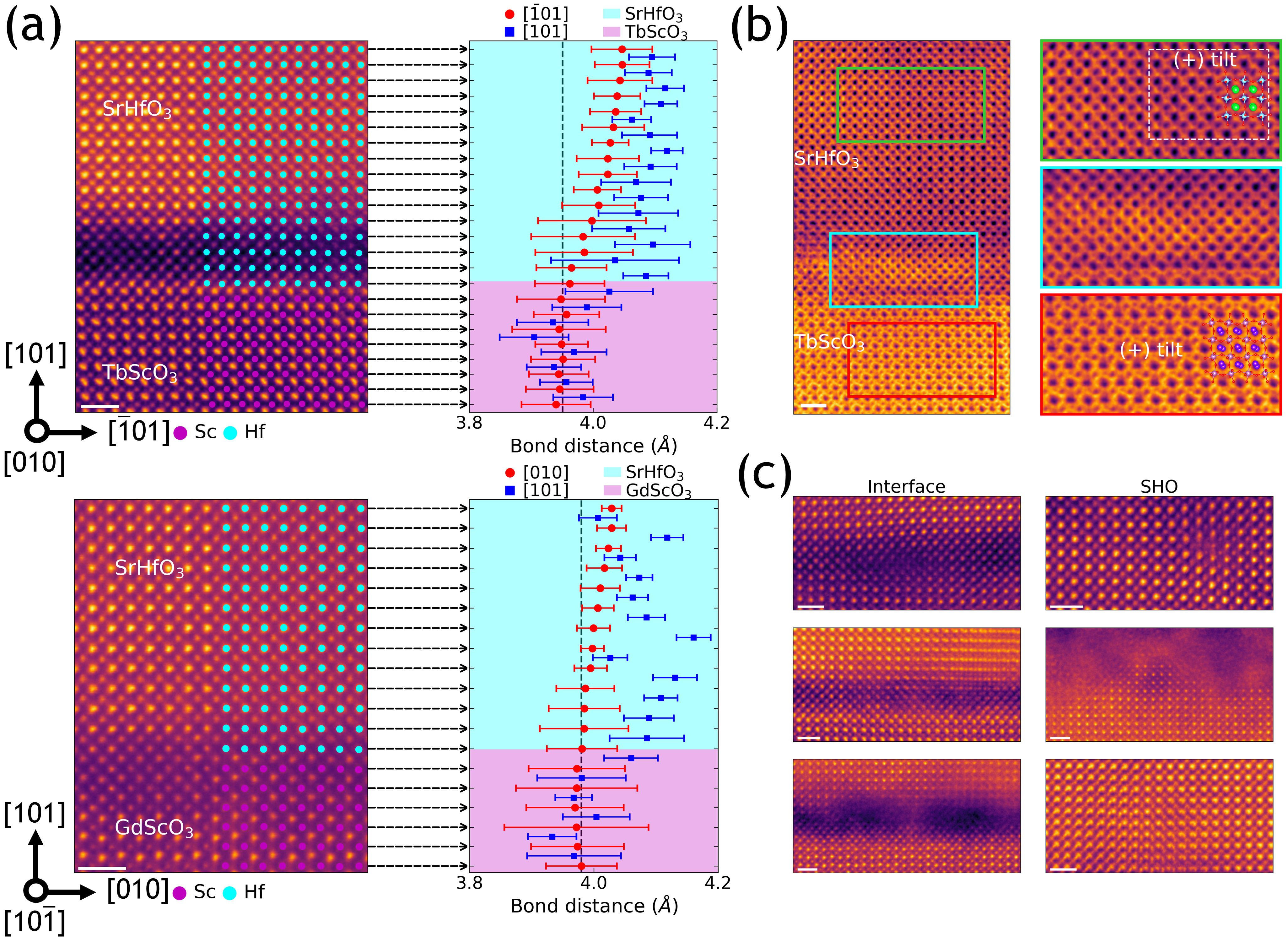}
            \caption{(a) Atomic resolution HAADF image showing the cross-section of the SHO-TSO (top) and SHO-GSO (bottom) with labeled B-site cations and the corresponding in-plane and out-of-plane bond distance profiles. The arrows mark the respective atomic-planes in the image and the bond-distance plot. The error bars in the bond distance profiles correspond to the standard deviation. The vertical dashed lines mark the in-plane and out-of-plane Sc-Sc bond distances for TSO and GSO respectively. The crystallographic orientations correspond to that of the TSO and GSO substrates. (b) Atomic-resolution ABF image (left) for an SHO-TSO sample. The higher-resolution ABF images (right) are extracted from the regions marked in same colors as the borders on the left ABF image. (c) Atomic-resolution HAADF images showing defects at the SHO-TSO/GSO interfaces and in the bulk of the SHO films. Scale bars in (a), (b) and (c) correspond to 1~nm.}
            \label{fig:STEM-atomic-res}
        \end{figure}

        The diffraction patterns for the P4mm phase are also within the measurement limits. However, based on the SHG data shown previously, we can rule out the presence of the non-centrosymmetric P4mm phase. Additionally, we also observe non-orthogonality in CBED patterns between the in-plane and out-of-plane Bragg vectors as well as some very weak reflections, which correspond to small domains with Pnma phase in the SHO films, which may be due to strain relaxation and other structural defects in the material. This phase impurity is also consistent with atomic-resolution imaging described later in Figure~\ref{fig:STEM-atomic-res}.

        To better understand the microstructure and epitaxy of the SHO films, we have further calculated strain for the SHO film using 4D-STEM CBED data. The strain maps for SHO-GSO[10$\bar1$] and TSO [010] are shown in Figures~\ref{fig:4D-STEM}(d) and~\ref{fig:4D-STEM}(h) respectively. The green circles in Figures \ref{fig:4D-STEM}(b) and \ref{fig:4D-STEM}(f) mark the Bragg disks selected for in-plane and out-of-plane strain calculation. Given that SHO has a larger unit cell volume than the GSO and TSO substrates, and the films are grown under a compressive in-plane strain, $\approx$-3\%, we expect a high out-of-plane tensile strain. This can be clearly observed in the out-of-plane strain maps for SHO-GSO and SHO-TSO as shown in Figures~\ref{fig:4D-STEM}(d) and \ref{fig:4D-STEM}(h). The strain maps show non-uniform strain for the in-plane lattice directions for both SHO-GSO and SHO-TSO samples. For a perfect epitaxy, the strain along the in-plane direction should be minimal. However, we find that for both samples, there are regions that correspond to good epitaxy, but a large fraction of the SHO film shows various degrees of tensile strain along the in-plane direction. The observation of tensile strain along the in-plane direction means that some interfacial strain relaxation has occurred due to misfit dislocations, which can be observed in HAADF measurements as well. This non-uniformity in strain maps can also be observed in LAADF imaging mode. The LAADF imaging mode is sensitive to strain contrast, where non-uniform image contrast reveals non-uniform epitaxy and a high density of defects, as shown in Figure S15. Therefore, the non-uniform strain in the SHO films can be directly attributed to phase impurities and a high density of defects arising from the epitaxial mismatch between the film and substrates.

        We have further performed atomic-resolution HAADF and ABF imaging to identify the changes in bond distances (Figure~\ref{fig:STEM-atomic-res}(a)), BO\textsubscript{6} octahedral tilt patterns (Figure~\ref{fig:STEM-atomic-res}(b)) and structural defects (Figure~\ref{fig:STEM-atomic-res}(c)) for the SHO-TSO and SHO-GSO samples. Figure~\ref{fig:STEM-atomic-res}(a) shows atomic-resolution HAADF images for the SHO-TSO (top) and SHO-GSO (bottom) samples. Since, intensity in a HAADF image is approximately proportional to the squared atomic number (\~Z\textsuperscript{2}) of the atomic column~\cite{Z-contrast} each atomic column in the substrate and the film can be clearly distinguished. For the HAADF images shown in Figure~\ref{fig:STEM-atomic-res}(a), we have labeled the B-site cations (Sc (Z~=~21) for TSO/GSO and Hf (Z~=~72) for SHO). The HAADF images for the TSO and GSO substrates match well with the expected Pnma phase for the substrates along their respective crystallographic orientations. The atomic models for the TSO and GSO substrates are shown in Figure S2. Consistent with the substrate crystallographic orientation (as marked in Figure~\ref{fig:STEM-atomic-res}(a)), we have calculated the in-plane and out-plane B-site cation bond distances (Sc-Sc for TSO/GSO and Hf-Hf for SHO) using the extracted Sc~and~Hf atomic positions from the corresponding SHO-TSO and SHO-GSO HAADF images. For the projections shown in Figure~\ref{fig:STEM-atomic-res}(a), the expected in-plane [$\bar1$01] and out-of-plane [101] Sc-Sc bond distance for TSO is 3.95~\AA. While for GSO, the expected in-plane [010] and out-of-plane [101] Sc-Sc bond distance is 3.98~\AA. These expected Sc-Sc bond distances are marked as vertical dashed lines in the bond distance profiles (Figure~\ref{fig:STEM-atomic-res}(a)). 
        
        For an ideal epitaxial relationship, the in-plane bond distances for the film (Hf-Hf) should be constrained to that of the substrate (Sc-Sc) for the entirety of the film thickness, while the out-of-plane bond distance should undergo relaxation (tensile for SHO) as we move further away from the substrate. As shown in Figure~\ref{fig:STEM-atomic-res}(a), we observe that the calculated in-plane and out-of-plane Sc-Sc bond distances show a good match for both TSO and GSO. We find that the out-of-plane Hf-Hf bond distance are larger compared to that of Sc-Sc, which is expected for the SHO film because of the application of compressive strain by both TSO and GSO substrates. However, for both SHO-TSO and SHO-GSO, we observe that the Hf-Hf in-plane bond distance is not constrained to that of the substrate, i.e. the Sc-Sc in-plane bond distance. This continued relaxation of the in-plane bond distances as we move away from the substrate results in the inability to achieve phase control for the SHO film growth. Moreover, the lack of epitaxial growth constraints results in the growth of SHO films that have a non-uniform interface with a high density of defects (as shown in Figures~\ref{fig:STEM-atomic-res}(c) and~S15. We also observe higher error bars in the bond distance profiles close to the interface, which is caused by chemical diffusion at the interface resulting in the formation of an intermixed layer, as also revealed by STEM-EELS data in Figure~\ref{fig:STEM-EELS}. The diffused interface along with the high-density of defects results in non-uniform strain accommodation as shown by atomic-resolution imaging (Figure~\ref{fig:STEM-atomic-res}(a)) as well as 4D-STEM analysis (Figure~\ref{fig:4D-STEM}).
    
        As is the case with most perovskite structures, the BO\textsubscript{6} octahedra can exhibit a variety of tilt patterns, which forms a basis of polymorphism in SHO. To ascertain the HfO\textsubscript{6} octahedral tilt patterns, we have performed atomic-resolution ABF imaging as shown in Figure~\ref{fig:STEM-atomic-res}(b). We observe that for the TSO substrate, ScO\textsubscript{6} octahedra show the expected (+) tilt pattern along [010] consistent with the TSO Pnma phase. However, for the SHO film, we observe non-uniform octahedral tilt patterns, where some regions exhibit (+) tilt patterns, as shown in Figure~\ref{fig:STEM-atomic-res}(b), which corresponds to a Pnma phase. While some regions don’t show any octahedral tilting, which corresponds to pseudocubic [100] and [010] orientations of I4/mcm and P4/mbm phases. Consistent with STEM-EELS data as shown Figure~\ref{fig:STEM-EELS}, we are unable to resolve the oxygen sublattice in the SHO film at the interface, because of a high concentration of oxygen vacancies at the SHO-TSO interface. Moreover, apart from a chemically diffused interface, we also observe a high density of structural defects at the SHO-TSO/GSO interface as shown in Figure~\ref{fig:STEM-atomic-res}(c). These defects include dislocations, stacking faults, misorientations and polycrystalline domains. Additionally, we also observe amorphous domains at the interface as well as at the SHO film surface. This confirms that it is unfeasible for the SHO films to accommodate a high compressive strain beyond a few unit cells. Figure~S15 shows wide field-of-view HAADF and LAADF images, where non-uniform strain contrast is evident across the cross-section of the SHO-films. This further illustrates non-uniform epitaxy and a high density of defects in the SHO films.
    
        Overall, using a combination of STEM-EELS, 4D-STEM CBED, and atomic-resolution imaging, we show that hMBE can yield highly crystalline SHO films on the TSO and GSO substrates. However, a high compressive strain results in a diffused interface with a high concentration of oxygen vacancies and structural defects, which makes phase control of the SHO film extremely challenging.

 \section{Discussion}
        
        The precise phase control of SHO films using hMBE is extremely challenging because of: (1) the various energetically competing SHO polymorphs, (2) high-temperature growth conditions, and (3) a large $\approx$-3\% compressive strain. As described earlier, the perovskite framework allows for the cooperative tilting of HfO\textsubscript{6} octahedra, which means that these polymorphs can readily undergo first and/or second-order phase transitions \cite{howard2005structures}. As a result, phase separation in SHO is likely unavoidable. The ground state phase for bulk SHO crystals at room-temperature is observed to be orthorhombic with a space-group symmetry of Pnma, which can undergo phase-transitions at higher temperatures \cite{High_Temp_Phases}. The Pnma orthorhombic structure undergoes phase-transition to a Cmcm orthorhombic structure between 400\textdegree C to 600\textdegree C. Further phase-transition to the I4/mcm tetragonal structure is observed between 600\textdegree C and 750\textdegree C. The I4/mcm tetragonal structure is thermodynamically stable till 1080\textdegree C, beyond which it undergoes phase-transition to a cubic Pm$\bar3$m phase. DFT calculations predict the Pnma phase to be thermodynamically favorable than the I4/mcm phase by 31 meV/fu at -3\% compressive strain, as shown in Figure \ref{fig:E vs Strain}. However, the SHO films were grown at a temperature of 1000\textdegree C, followed by cooling down to 400\textdegree C over 400 seconds, which makes them conducive to adopting the I4/mcm tetragonal structure. Further optimization of growth conditions, such as deposition temperature and prolonged annealing at different temperature ranges, could potentially be a systematic approach to achieve phase control in SHO films grown by hMBE. 

        We observe that the SHO film growth is initially pseudomorphic to the TSO and GSO substrates, as shown in Figure \ref{fig:STEM-atomic-res}. However, such a high compressive strain is eventually unsustainable, resulting in strain relaxation via the formation of a high density of planar defects at the interface, such as dislocations, stacking faults, and polycrystalline domains. The growth of a graded buffer layer could be a potential solution to the high compressive strain issue. A graded buffer layer, with intermediate in-plane lattice parameters in between the substrate and SHO, could help distribute strain gradually. However, the compositional complexity of ternary oxides and the refractory nature of some of the metals makes it challenging to identify a suitable buffer layer.
        
        Another critical aspect of phase control of SHO films using hMBE is the quality of substrate, in particular surface reduction of the substrate prior to growth. As shown by STEM-EELS (Figure \ref{fig:STEM-EELS}), the oxygen deficiency at the interface can be explained by the likely reduction of the TSO and GSO before the growth of SHO films. Although the TSO and GSO substrates were heated to 1000\textdegree C for 1hr in oxygen plasma to avoid surface reduction, the ultra-high vacuum and a high substrate temperature during hMBE could still result in the deleterious reduction of the substrate. This makes the initial $\approx$-2-4 u.c. of the SHO layer oxygen deficient, rendering the subsequent growth of the film suboptimal. The development of additional heat and chemical treatment methods for substrate prior to film growth could potentially alleviate the problem of surface reduction and will be the subject of future work. Recent developments in laser heating of substrates \cite{hensling2024state} suggest that this approach may be viable to improve the film-substrate interface during growth at temperatures of 1000 \textdegree C. 

\section{Conclusion}

We have successfully demonstrated the epitaxial growth of SHO films grown on GSO and TSO substrates using the TEMAH metal organic precursor in hMBE. We have achieved the first synthesis of SHO with -3\% biaxial in-plane strain, as confirmed by RSM measurements. The wide bandgap and insulating properties of the material make SHO an ideal material for capping layers and dielectric barriers in perovskite oxides films and heterostructures. Previous DFT studies predicted that SHO films would exhibit the polar P4mm phase when synthesized under compressive strain \cite{Piezoelectric, Bad_P4mm}. However, SHG measurements identified no SHG response, indicating the samples are centrosymmetric and ruling out the presence of the P4mm phase. The absence of polar distortions is in agreement with previous computational studies \cite{DFT,DFT_SEO} and synthesis experiments~\cite{PHO}. However, this contradicts one previous study that found SHO-STO films exhibited the P4mm phase~\cite{Bad_P4mm}, despite having less experimentally-determined compressive strain than the films synthesized in this work.

While DFT modeling showed a small range of structural energies between five possible phases of SHO, XAS measurements and 4D-STEM mapping of the samples rule out the presence of the Pnma, Pm$\bar3$m, and P4mm phases. The absence of the Pnma phase, which SHO, GSO, and TSO take in the bulk, supports previous first principal studies that predicted compressive strain could change the octahedral tilt pattern in SHO~\cite{DFT}. However, the absence of the cubic Pm$\bar3$m phase contrasts with a synthesis study which found SHO-STO films took on a cubic phase~\cite{PHO}. 4D-STEM measurements showed that the SHO-GSO and SHO-TSO films exhibit either I4/mcm and P4/mbm phase. Based on the DFT modeling presented, we conclude that the films most likely take on the I4/mcm structure rather than the P4/mbm due the 31~meV/fu difference in energy between the two structures. These results demonstrate that compressive strain tends to drive changes in octahedral tilt patterns in SrHfO$_3$ rather than polar distortions. Future works could examine the temperature dependence of the octahedral tilt patterns or explore ways to synthesize heterostructures or superlattices that drive SrHfO$_3$ into a different phase.

\begin{acknowledgement}

P.T.G. gratefully acknowledge support for synthesis and characterization from the National Science Foundation (NSF) under Award No. DMR-2045993. R.B.C. acknowledges support from the Air Force Office of Scientific Research for the film synthesis and characterization under Award No. FA9550-20-1-0034. W.J. acknowledges support from the US Department of Energy (DOE) Office of Science under DE-SC0023478 and C.T. acknowledges support by NSF EPM Grant No. DMR-212987. P.T.G. and R.B.C. acknowledge the Auburn University Easley Cluster for support of this work. B.K. gratefully acknowledges the usage of Stampede2 and Stampede3 at TACC through allocation DMR-110093 from the Advanced Cyberinfrastructure Coordination Ecosystem: Services \& Support (ACCESS) program, which is supported by National Science Foundation grants \#2138259, \#2138286, \#2138307, \#2137603, and \#2138296. XRD measurements were performed using an instrument acquired through the NSF Major Research Instrumentation Program under Award No. DMR-2018794. This research used resources of the Advanced Photon Source, a U.S. Department of Energy (DOE) Office of Science user facility operated for the DOE Office of Science by Argonne National Laboratory under Contract No. DE-AC02-06CH11357. A.S.T and R.F.K. were supported by the Office of Basic Energy Sciences, U.S. Department of Energy, award number DE-SC0025396. Acquisition of the UIC JEOL ARM200CF was supported by an MRI-R\textsuperscript{2} grant from the National Science Foundation (DMR-0959470). The Gatan Continuum GIF acquisition at UIC was supported by an MRI grant from the National Science Foundation (DMR-1626065).

\end{acknowledgement}

\begin{suppinfo}

Additional DFT models assuming other elastic responses in SrHfO$_3$; additional STEM, XAS, XPS, and XRD data; atomic models of various SrHfO$_3$ space groups, along with GdScO$_3$ and TbScO$_3$ structures. 

\end{suppinfo}

\bibliography{references_GR.bib}

\end{document}